\documentclass[aps,PRB,reprint,linenumbers,preprintnumbers]{revtex4-1}
\usepackage[T1]{fontenc}
\usepackage{times}
\usepackage{lettrine}
\usepackage{graphicx,color}
\usepackage{amsfonts,amsmath,amssymb,amsbsy}
\usepackage{fancyhdr}
\usepackage{float}
\usepackage{placeins} 

\definecolor{orange}{rgb}{1,0.41,0.13}
\usepackage[colorlinks=true, linkcolor=blue, citecolor=blue, urlcolor=blue]{hyperref}
\def\section#1{\medskip\noindent\textbf{\large{#1}}\par}
\makeatletter
\renewcommand{\fnum@figure}{\figurename~\textbf{\thefigure}}
\makeatother
\renewcommand{\figurename}{\textbf{Fig.}}

\let\mathbf=\boldsymbol
\let\Gamma=\varGamma
\let\Upsilon=\varUpsilon
\def\emph#1{\textcolor{red}{#1}}

\begin{document}
\nolinenumbers
\title{Helical Magnon Frequency Comb in Synthetic Antiferromagnetic Skyrmion Lattices}

\author{Xuejuan Liu}
\thanks{These authors contributed equally to this work.}
\affiliation{Shenzhen Key Laboratory of Ultraintense Laser and Advanced Material Technology, Center for Intense Laser Application Technology, and College of Engineering Physics, Shenzhen Technology University, Shenzhen 518118, China}
\affiliation{College of Applied Sciences, Shenzhen University, Shenzhen 518060, China}
\author{Xingen Zheng}
\thanks{These authors contributed equally to this work.}
\affiliation{The Center for Advanced Quantum Studies and School of Physics and Astronomy, Beijing Normal University, Beijing 100875, China}
\affiliation{ Key Laboratory of Multiscale Spin Physics, Beijing Normal University, Beijing 100875, China}

\author{Zhixiong Li}
\affiliation{School of Physics, Central South University, Changsha 410083, China}

\author{Xiaoguang Li}
\affiliation{Shenzhen Key Laboratory of Ultraintense Laser and Advanced Material Technology, Center for Intense Laser Application Technology, and College of Engineering Physics, Shenzhen Technology University, Shenzhen 518118, China}
\author{Cangtao Zhou}
\affiliation{Shenzhen Key Laboratory of Ultraintense Laser and Advanced Material Technology, Center for Intense Laser Application Technology, and College of Engineering Physics, Shenzhen Technology University, Shenzhen 518118, China}
\author{Hui Li}
\affiliation{Shenzhen Key Laboratory of Ultraintense Laser and Advanced Material Technology, Center for Intense Laser Application Technology, and College of Engineering Physics, Shenzhen Technology University, Shenzhen 518118, China}
\author{Haipeng Sun}
\affiliation{Shenzhen Key Laboratory of Ultraintense Laser and Advanced Material Technology, Center for Intense Laser Application Technology, and College of Engineering Physics, Shenzhen Technology University, Shenzhen 518118, China}

\author{Peng Yan}
\email[Email:~]{yan@uestc.edu.cn}
\affiliation{School of Physics and State Key Laboratory of Electronic Thin Films and Integrated Devices, University of Electronic Science and Technology of China, Chengdu 610054, China}

\begin{abstract}\noindent
Synthetic antiferromagnetic skyrmion lattices (SAF-SkLs), consisting of two ferromagnetic SkLs coupled antiferromagnetically with compensated magnetization, provide a promising platform for robust nonlinear magnonics. Here, we investigate helical magnon frequency comb (HMFC) generation in a SAF-SkL by combining analytical modeling with micromagnetic simulations. We show that nonlinear coupling between helical magnon edge states and the skyrmion gyration produces frequency combs with pronounced edge localization. The HMFC exhibits strong enhancement exclusively when the driving frequency lies within the helical edge-state band, whereas the interior response remains negligible. This frequency selectivity confirms the essential role of helical edge modes in localized nonlinear frequency conversion. We further demonstrate that the interlayer antiferromagnetic coupling reconstructs the magnon spectrum, thereby tuning the comb spacing and the number of comb teeth while also redistributing modal energy. Our results establish SAF-SkLs as a tunable platform for edge-localized HMFCs and suggest a route toward robust coherent magnonic signal processing.
\end{abstract}

\date{\today}
\preprint{}

\maketitle

\section{Introduction}
\label{se:Introduction}

\begin{figure*}[!t]
	\centerline{\includegraphics[width=0.7\textwidth]{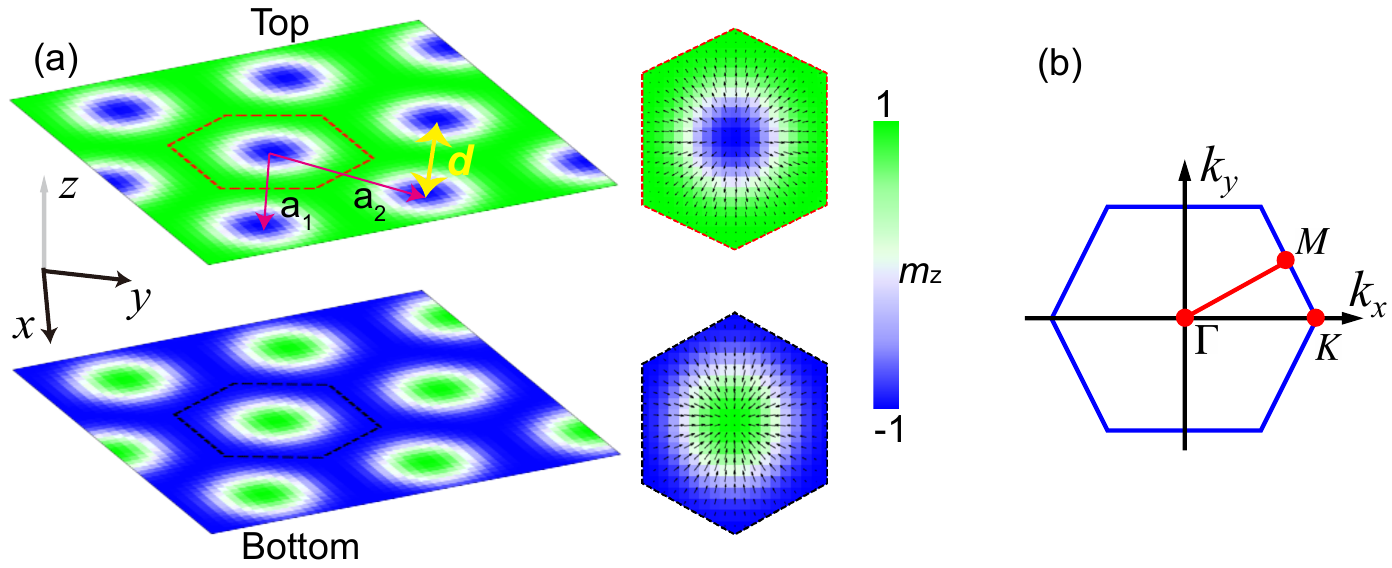}}
	\caption{ \textbf{Schematic of the Bilayer SAF-SkL texture.} \textbf{a} The real-space SAF-SkL structure with lattice constant $d=40$ nm. The right panels show enlarged views of the magnetic unit cell in the two layers, as indicated by the dashed hexagons in the left panel. The arrows indicate the spin vectors. \textbf{b} First Brillouin zone of a single SkL, with high-symmetry points $\Gamma$, $M$, and $K$ labeled.}
	\label{FIG1}
\end{figure*}
Magnon frequency combs (MFCs) have emerged as a versatile platform for precision metrology, quantum information processing, and on-chip signal manipulation, owing to their capability of coherent frequency conversion without optoelectronic interfaces and their intrinsically high spectral resolution \cite{Heins2025,Yuan2022}. In the microwave regime, a variety of mechanisms have been proposed and demonstrated for MFC generation, including three-magnon scattering in nonuniform magnetic textures such as skyrmions \cite{Wang2021}, vortices \cite{Wang2022PRL}, bimerons \cite{Zhang2024}, and domain walls \cite{zhou2021JMMM}, resonance-enhanced magnetostriction \cite{Liu2023}, magnetostrictive mechanical \cite{MaFS2026NC} and four-magnon processes in low-dimensional nanowires \cite{Hula2022}. In addition, photoinduced nonlinearities, exceptional-point engineering \cite{Wang2024}, and Floquet-state-driven vortex interactions \cite{Heins2026APL} have further expanded the accessible parameter space for comb formation. Despite this rapid progress, existing approaches are inherently susceptible to material imperfections and environmental fluctuations, which limit coherence times and frequency stability and thus hinder their applicability in quantum-grade technologies. Achieving robust MFCs with intrinsic protection against disorder therefore remains an outstanding challenge.

Topological magnonics provides a natural route to address this issue. Systems with topologically protected chiral edge states exhibit magnon transport that is immune to backscattering from defects, enabling robust linear functionalities such as diodes and beam splitters \cite{Mook2021,Wang2017}. Extending this robustness to the nonlinear regime is highly desirable but remains largely unexplored. Two-dimensional skyrmion lattices (SkLs) offer a particularly attractive platform: their nontrivial spin textures generate emergent gauge fields and endow magnon excitations with topological protection, enabling controllable magnon transport and nonlinear interactions \cite{Weber2022,Roldan2016,Ma2015, Liu2024}. Recently, a topological magnon frequency comb (TMFC) was demonstrated in a ferromagnetic SkL \cite{Li2025}, establishing the feasibility of combining topology with nonlinear frequency conversion. However, the finite net magnetization of ferromagnets inevitably induces dipolar interactions and stray fields, which can distort the magnetic texture and reduce its robustness.

These limitations could be circumvented by utilizing skyrmions in antiferromagnets
\cite{Chen2025,Shen2020,Diaz2019,Barker2016,Xia2021}. In these materials, antiferromagnetically coupled skyrmions reside in different sublattices and have opposite core polarities and thus opposite topological charges. Antiferromagnetic skyrmions have been stabilized in a variety of systems, including bulk MnSc$_{2}$S$_{4}$ \cite{Gao2020}, Fe$_{2}$O$_{3}$ \cite{Jani2021}, exchange-biased IrMn antiferromagnets \cite{Rana2021} , and synthetic antiferromagnets \cite{Juge2022}. Among these, compensated synthetic antiferromagnets combine key advantages for investigating the dynamics of antiferromagnetic skyrmions and their utilization in devices. SAFs consist of ferromagnetic layers separated by nonmagnetic spacers and are antiferromagnetically coupled via the Ruderman-Kittel-Kasuya-Yosida interlayer interaction \cite{Parkin1990, Pham2024}. 

Here, we demonstrate the generation of helical magnon frequency combs (HMFCs) in a synthetic antiferromagnetic skyrmion lattice (SAF-SkL). 
By calculating the magnon band structure of the SAF-SkL, we identify counterpropagating helical edge states arising from the antiferromagnetically coupled skyrmion layers. Micromagnetic simulations reveal that these helical edge states assist nonlinear frequency-comb generation through a three-magnon interaction with the gyrotropic mode of skyrmions. The resulting HMFC exhibits pronounced edge localization, and its comb spacing is set by skyrmion's gyration. We further show that the interlayer coupling provides an effective control parameter for tuning the comb spacing, the number of comb teeth, and the redistribution of modal spectral weight. These results establish the SAF-SkL as a tunable platform for edge-localized nonlinear magnon dynamics and suggest a route toward robust nonlinear magnonics.

\vbox{}
\section{Results and discussion}
\label{se:Results}

\noindent
\textbf{Theoretical model.} 

We theoretically calculate the magnon band structure and identify helical magnon edge modes emerging in a system with zero net magnetization, which are corroborated by micromagnetic simulations. We report two distinct types of MFCs: a low-frequency MFC generated by coupling between the  hybrid mode and the collective gyrotropic mode, and an HMFC arising from interactions between helical magnons and the gyrotropic mode. The HMFC exhibits counterclockwise propagation in the top layer and clockwise propagation in the bottom layer, revealing its helical nature. 

We consider a bilayer synthetic antiferromagnetic system hosting a two-dimensional triangular N\'{e}el-type SkL, stabilized by the interfacial Dzyaloshinskii-Moriya interaction (DMI), as illustrated in Fig. \ref{FIG1}a. The lattice basis vectors are $\textbf{a}_{1}=d\hat{\textbf{x}}$ and $\textbf{a}_{2}=\frac{d}{2}\hat{\textbf{x}}+\frac{\sqrt{3}d}{2}\hat{\textbf{y}}$, where $d$ is the distance between nearest-neighbor skyrmions. The corresponding first Brillouin zone of a single SkL is shown in Fig. \ref{FIG1}b, with high-symmetry points $\Gamma$, $M$, and $K$. 
   The system is described by the spin Hamiltonian
\begin{eqnarray}
 \begin{aligned}
{\cal H}=&-\sum_{<i,j>}J_\mathrm{ex}\mathbf{S}_{i}\cdot \mathbf{S}_{j}-\sum_{<i,i^{'}>}J_\mathrm{af} \mathbf{S}_{i}\cdot \mathbf{S}_{i'}\ +\\&
    \sum_{<i,j>}D(\hat{z}\times \mathbf{\hat{r}}_{ij})\cdot (\mathbf{S}_{i}\times \mathbf{S}_{j})-K\sum_{i}(S^{z}_{i})^{2},
  \end{aligned}
\label{eq:1}
\end{eqnarray}
where $\mathbf{S}_{i}$ $(\mathbf{S}_{j})$ is the spin-$S$ operator at site  $i$ ($j$). Here, $\langle i,j \rangle$ runs over nearest-neighbor sites within each layer, whereas $\langle i,i' \rangle$ denotes vertically aligned sites in the two layers. $J_\mathrm{ex}$ is the intralayer exchange interaction; $J_\mathrm{af}$ is the interlayer antiferromagnetic coupling; $D$ is the interfacial DMI; and $K$ is the perpendicular uniaxial anisotropy. MuMax3 is used for the later simulations (see Methods), and the following material parameters are adopted \cite{SampaioNN2013}: the saturation magnetization  $M_\mathrm{s} = 0.58 \times 10^6$ A/m, the intralayer exchange stiffness $A_\mathrm{ex} = 1.5 \times 10^{-11}$ J/m, the interlayer antiferromagnetic exchange stiffness $A_\mathrm{af} = -1.5 \times 10^{-12}$ J/m, the interfacial DMI constant $D_\mathrm{ind} = 3$ mJ/m$^2$, the perpendicular magnetic anisotropy $K_\mathrm{u} = 0.36 \times 10^6$ J/m$^3$, and Gilbert damping coefficient $\alpha = 0.001$. These micromagnetic parameters are mapped onto the Hamiltonian parameters in  Eq. \eqref{eq:1} via the spin quantum number $S=M_\mathrm{s}a^{3}/g\mu_{B}$ with $g$-factor $g = 2$, $J_\mathrm{ex} = 2A_\mathrm{ex}a/S^{2}$, $J_\mathrm{af} = 2A_\mathrm{af}a/S^{2}$, $D = D_\mathrm{ind} a^{2}/S^{2}$, $K = K_{u}a^{3}/S^{2}$ \cite{Wang2026PRB}.

\begin{figure*}[!t]
	\centerline{\includegraphics[width=0.75\textwidth]{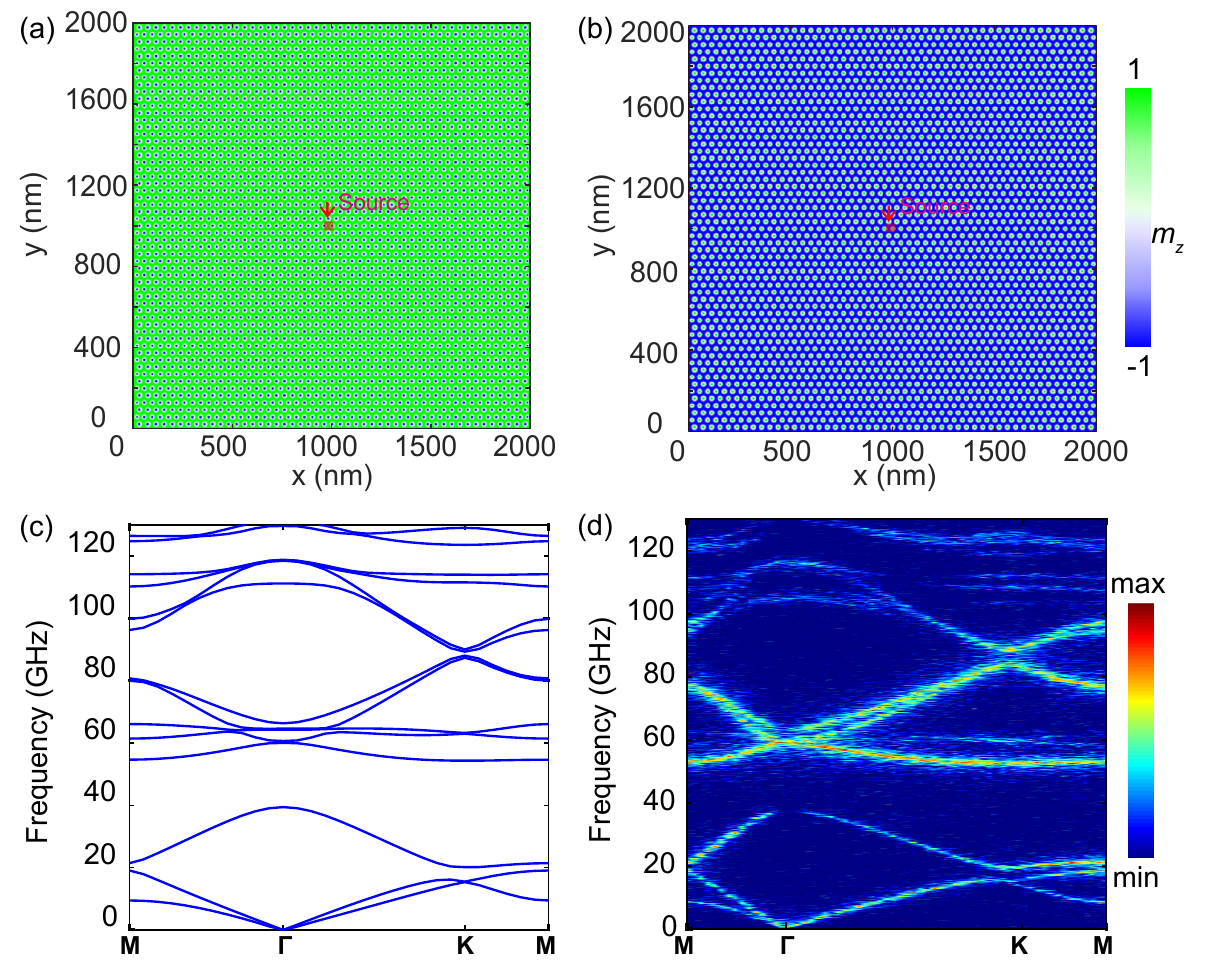}}
	\caption{\textbf{Bulk magnon band structure of the SAF-SkL.} \textbf{a}, \textbf{b} Schematics of the top and bottom SkLs, with a localized excitation magnetic field applied to a central quadrilateral region (black shaded area). \textbf{c} Theoretical bulk magnon band structure of the SAF-SkL along the $M-\Gamma-K-M$ path. Only the lowest fourteen bands are shown. \textbf{d} Micromagnetically simulated band structure of the same system along the identical path.}
	\label{FIG2}
\end{figure*}

\begin{figure*}[!t]
	\centerline{\includegraphics[width=1\textwidth]{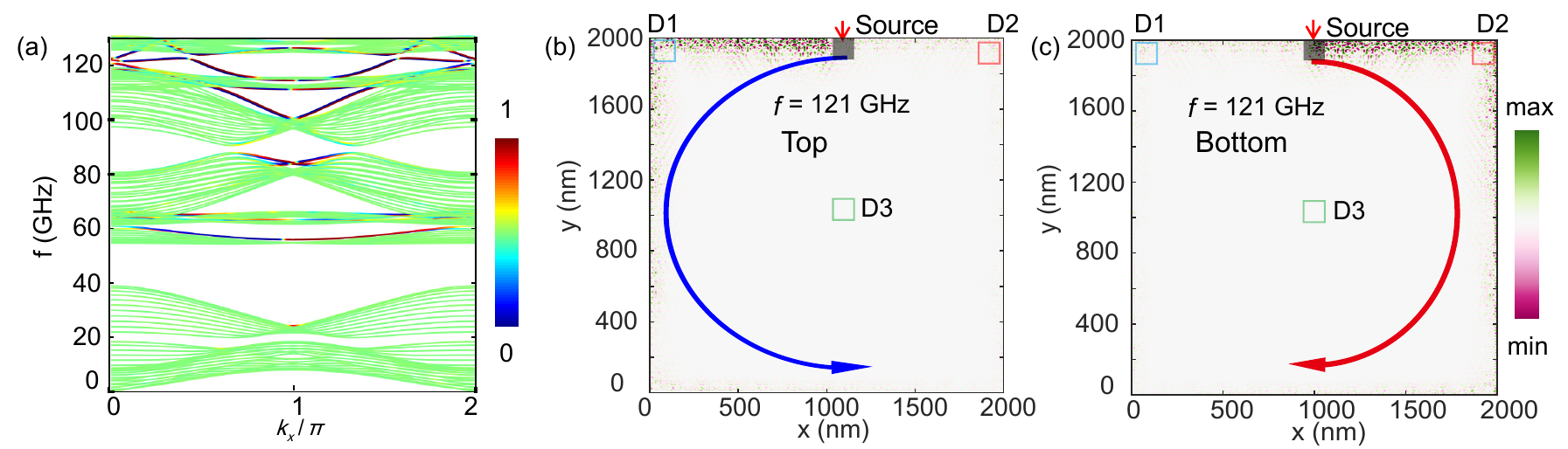}}
	\caption{\textbf{Helical edge states with opposite chirality in the bilayer SAF-SkL. } \textbf{a} Strip magnon spectrum colored by the normalized transverse position $<y>/L_y = \sum_{\mu} (y_{\mu} - y_{\rm min}) \rho(\mu)/L_y$. Here, $\mu$ labels the spin sites in the strip unit cell; $y_{\mu}$ is the $y$ coordination of the $\mu$-th cell; $L_y = y_{\rm max} - y_{\rm min}$; $\rho(\mu) = |u_{\mu}|^2 + |v_{\mu}|^2$ is the local magnon weight of the Bogoliubov-de Gennes (BdG) eigenmode $\psi = (u, v)^T$. The value $<y>/L_y$ is from 0 to 1 and indicates edge localization. \textbf{b}, \textbf{c} Real-space snapshots at 121 GHz with color-coded intensity, showing counterpropagating edge transport in the top (b) and bottom (c) layers (arrows indicate propagation directions).}
	\label{FIG3}
\end{figure*}

\vbox{}\noindent
\textbf{Helical magnon edge state in SAF-SkLs.}

A stable SAF-SkL with an equilibrium nearest-neighbor skyrmion spacing of $d=40$ nm is first realized. The sample has lateral dimensions of $2 \times 2$  $\mu$m$^2$ and a thickness of 4 nm, discretized into cubic cells of $2 \times 2 \times 2$ nm$^3$. The SAF-SkL consists of two coupled SkL layers, with the top and bottom layers illustrated in Figs. \ref{FIG2}a and b. Based on this relaxed magnetic texture, we obtain the low-energy magnon band structure and corresponding eigenstates within the Holstein-Primakoff formalism combined with para-unitary diagonalization [see Sec. 1 in Supplementary Materials (SM)]. Figure \ref{FIG2}(c) shows the lowest fourteen bands along the high-symmetry path $M-\Gamma-K-M$. In the absence of interlayer coupling, the spectrum reduces to two degenerate copies of the single-layer SkL bands \cite{Li2025}. Finite interlayer antiferromagnetic coupling lifts this degeneracy and reshapes the band structure. We confirm that the helical properties of the magnon bands in the SAF-SkL originate from the topological chiral edge states of the uncoupled limit \cite{xingen2026}. 

To verify these predictions, we perform simulations on an infinite two-dimensional SAF-SkL subject to periodic boundary conditions along the $x$ and $y$ directions [see methods for simulation details)]. As indicated by the black shaded regions in Figs. \ref{FIG2}a and b, the excitation is applied to a central $40 \times 40$ nm$^2$ region. The resulting bulk magnon bands of the SAF-SkL are shown in Fig. \ref{FIG2}d, which exhibit excellent agreement with theoretical predictions, confirming the key role of interlayer coupling.

 To reveal the helical nature of these bands, we calculate the spectrum using a ribbon geometry with open boundaries along the $y$ direction and periodic along the $x$ direction, as shown in Fig. \ref{FIG3}a. The gapped branches near the upper and lower edges (about 118--123.5 GHz) are clearly identified as edge modes. These edge states appear as nearly doubly degenerate counterpropagating branches, primarily associated with the two constituent layers. To visually distinguish these two branches, we apply a very small out-of-plane magnetic field (about 10 mT) in the strip calculation. This field slightly lifts the degeneracy between them but does not alter the qualitative structure of the spectrum. To probe the edge dynamics, sinusoidal microwave sources are applied at the $y$-direction boundaries [black shaded regions in Figs. \ref{FIG3}b and c]. The sources simultaneously excite both magnetic layers, with open boundaries imposed in both $x$ and $y$ directions. The subsequent time evolution is monitored, and representative snapshots of magnon propagation are presented in Figs. \ref{FIG3}b and c. The simulated dynamics reveal a helical transport mode, in which edge excitations in the top and bottom layers propagate with opposite chirality. The bilayer system thus hosts counterpropagating edge channels, consistent with the band structure predictions. Furthermore, we examined the impact of defects on the edge states. As demonstrated in Fig. S1 [see Sec. 2 of the SM], the helical nature of the edge states persists even in the presence of defects, underscoring the remarkable robustness of the system.

\vbox{}\noindent
\textbf{Helical magnon frequency comb.}  
The helical nature of these robust edge states makes SAF-SkL an ideal platform for exploring nonlinear dynamics. A sinusoidal excitation field $\mathbf{h}(t) = h \, \sin(2\pi f_{\mathrm{l}} t) \, \hat{\mathbf{x}}$
with $f_{\mathrm{l}} = 121$ GHz and the amplitude $h$ varying from $0$ to $60$~mT in steps of $5$~mT, is applied at the source sites, as indicated by the black shaded areas in Figs. \ref{FIG3}b and c. The blue, red, and green rectangles in these figures denote the regions where the magnon spectra are calculated, labeled D1, D2, and D3, respectively. As shown in Fig. S2 [see Sec. 3 of the SM], the HMFC emerges at a low driving amplitude of approximately 1 mT and maintains a fixed comb spacing across a wide range of excitation amplitudes, rendering it experimentally feasible with substantially reduced power consumption. This low-threshold behavior is consistent with a three-magnon splitting process \cite{Wang2021}, where a high-frequency magnon decays into a low-frequency gyrotropic mode and a secondary magnon, facilitating the formation of the frequency comb.

\begin{figure}[!t]
\centerline{\includegraphics[width=0.5\textwidth]{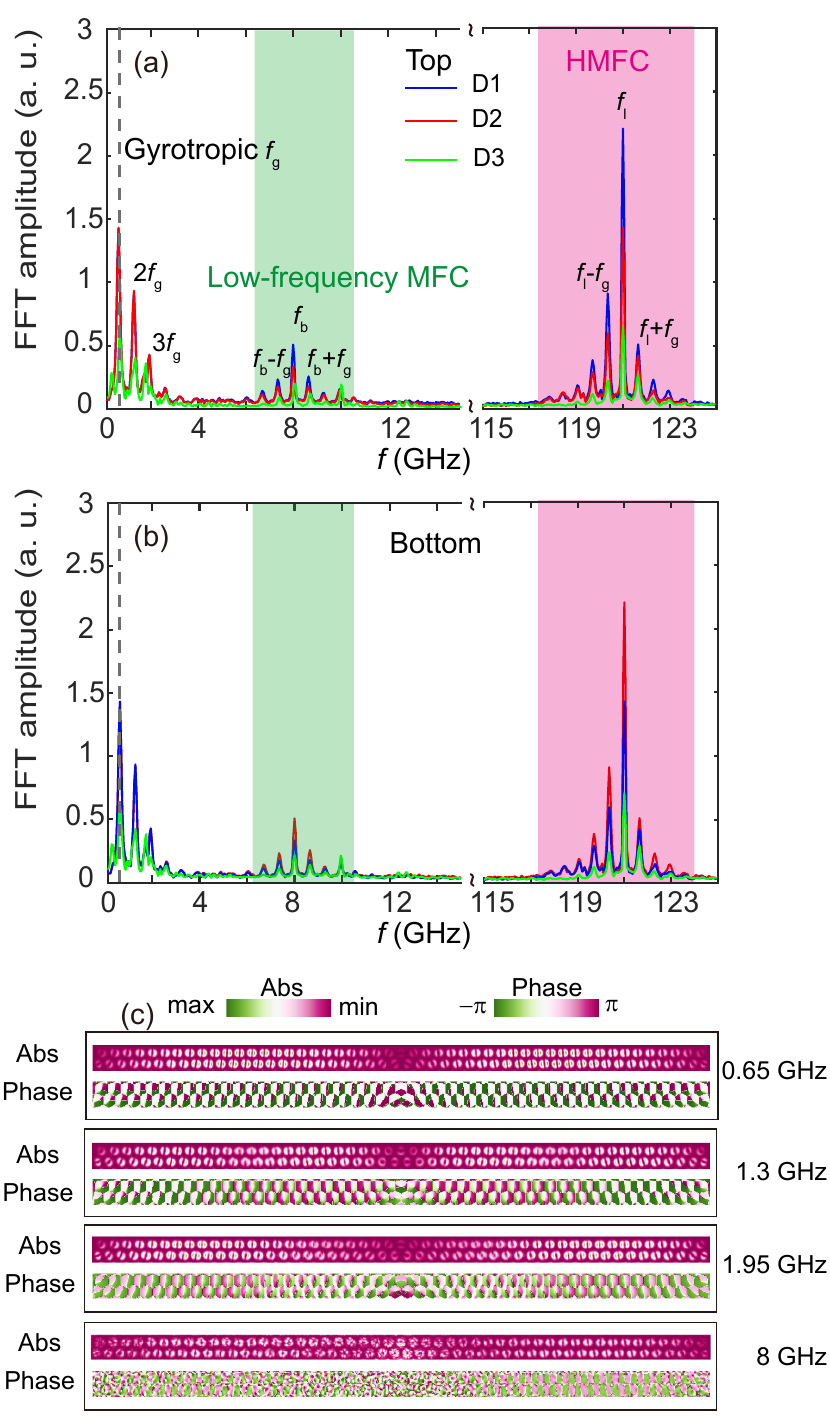}}
\caption{ \textbf{HMFC generation in the SAF-SkL.}  Spin wave spectra measured at positions D1, D2, and D3 in the top \textbf{a} and bottom \textbf{b} SkLs, respectively. \textbf{c} The spatial profiles of the spectral amplitudes (upper) and phases (lower) for several representative modes.}
\label{FIG4}
\end{figure}
The detailed spectra of the top and bottom layers under an excitation amplitude of 50 mT are presented in Figs. \ref{FIG4} a and b. Our analysis reveals that, for both the top and bottom SkLs, the spin-wave amplitude is sufficiently large to excite collective gyrotropic modes (including its high-order harmonics, $nf_\mathrm{g}$) and hybrid ($f_\mathrm{b}$) modes. The gyrotropic mode participates in three-magnon coupling with the applied spin wave and the hybrid mode, generating a low-frequency MFC \cite{Liu2024} and an HMFC with a frequency spacing exactly equal to 0.65 GHz. The high-frequency comb sidebands are prominent at the edge detectors D1 and D2 but strongly suppressed at the interior detector D3, demonstrating edge-localized nonlinear conversion. Further examination of the intensity distribution reveals that in the top SkL [Fig. \ref{FIG4}a], where spin waves propagate counterclockwise, the intensity at D1 exceeds that at D2, a difference attributable to dissipative decay during propagation. In contrast, in the bottom SkL [Fig. \ref{FIG4}b], where spin waves propagate clockwise, the intensity at D2 surpasses that at D1. These results confirm the helical origin of the observed MFCs. Furthermore, we restrict  spectral analysis to an 80$\times$2000 nm$^{2}$ region in the upper portion of the top-layer SkL. Figure \ref{FIG4}c displays the amplitude (upper) and phase (lower) at $f_\mathrm{g}$ (0.65 GHz), $2f_\mathrm{g}$ (1.3 GHz), $3f_\mathrm{g}$ (1.95 GHz), and $f_\mathrm{b}$ (8 GHz), and identifies the first three frequencies as the gyration mode and its harmonics, while the 8 GHz signal corresponds to a complex hybrid mode.

Prior investigations have revealed that inhomogeneous edge spin textures enhance nonlinear spin-wave mixing \cite{YuNC2025} and generate higher harmonics, but the resulting MFCs are achiral. In contrast, MFCs in a single ferromagnetic SkL are chiral, arising from four-magnon processes involving the dipolar field, with chirality originating from topological magnon bands \cite{Li2025}. Our HMFCs in SAF-SkL differ from both cases: they originate from three-magnon processes with uncanted edge magnetization \cite{Wang2021,LiuYPRB2024}. Therefore, the observed chirality is governed primarily by the helical edge channels rather than by edge canting.

\begin{figure*}[!htb]
\centerline{\includegraphics[width=0.8\textwidth]{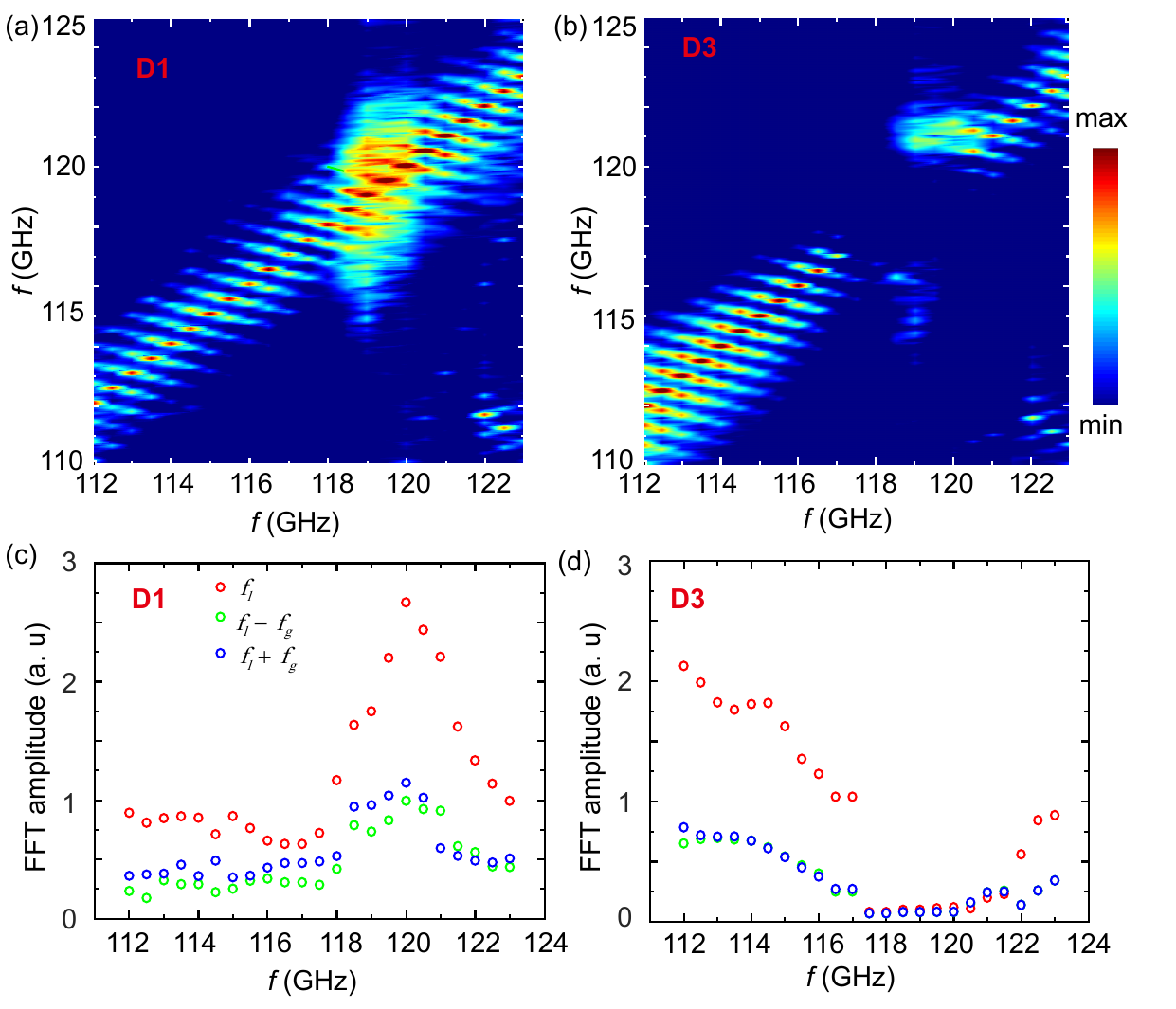}}
\caption{\textbf{Dependence of the HMFC on the driving frequency.}
\textbf{a}  Measured spin wave spectra in region D1. \textbf{b} Measured spectra in region D3. \textbf{c} Extracted amplitudes of the primary peak $f_\mathrm{l}$ and the sidebands $f_\mathrm{l}\pm f_\mathrm{g}$ as a function of the driving frequency in D1. \textbf{d} Corresponding amplitudes for D3. The driving amplitude is maintained at 50 mT. }
\label{FIG5}
\end{figure*}

We further investigate the dependence of the HMFC on the driving frequency. As shown in Fig. \ref{FIG5}a, a pronounced HMFC signal appears in region D1 only when the spin-wave frequency lies within the helical edge-state band. Over the same frequency range, region D3 exhibits negligible frequency comb activity [Fig. \ref{FIG5}b]. The  intensity profiles of the primary ($f_\mathrm{l}$), sum-frequency ($f_\mathrm{l}+f_\mathrm{g}$), and difference-frequency ($f_\mathrm{l}-f_\mathrm{g}$) peaks as functions of the driving frequency are presented in Figs. \ref{FIG5}c and d. These results indicate that the helical edge-state band provides the dominant channel for efficient HMFC generation under the present excitation geometry. The excitation efficiency at the edge far exceeds that in the interior, demonstrating that helical edge states locally enhance nonlinear frequency conversion.

The influence of interlayer coupling on MFCs has been investigated in various magnetic systems: it can tune the comb spacing in synthetic ferrimagnets \cite{LiuYPRB2024}, while its effect is negligible on isolated skyrmionium states in synthetic antiferromagnets \cite{GaoPRB2026}. Here, we systematically investigate its role in HMFCs.  Figure S3 [see details in Sec. 4 of SM] presents the MFC spectra measured in regions D1, D2, and D3 under three representative antiferromagnetic coupling strengths. Specifically, the helical magnon states originally centered around 121 GHz undergo a substantial frequency shift under strong coupling, leading to a redistribution of the excitation response from the edge-localized regions (D1 and D2) toward the central region (D3). These results demonstrate that interlayer coupling in SAF-SkLs not only governs the comb spacing and the number of comb teeth but more fundamentally triggers mode competition and energy redistribution through helical magnon band reconstruction. This mechanism is physically distinct from the behavior observed in isolated skyrmion or skyrmionium systems.


\section{Conclusions}
In summary, this study demonstrates helical magnon frequency combs in a SAF-SkL. Via nonlinear three-magnon interactions involving magnon edge states, HMFCs are generated with a fixed spacing determined solely by the gyrotropic frequency. The combs exhibit pronounced edge localization and remain robust over a broad range of driving amplitudes, while their efficiency is strongly enhanced within the helical edge-state band. Crucially, the interlayer coupling in SAF-SkLs not only controls the comb spacing and the tooth number, but more fundamentally drives a mode competition and energy redistribution through band reconstruction--a mechanism distinct from isolated skyrmion or skyrmionium systems. The coexistence of helical robustness and nonlinear tunability establishes a new paradigm for MFC generation, opening a route toward low-dissipation, highly stable magnonic devices for coherent signal processing and quantum technologies.

\section{Methods}
\label{se:Method}
\noindent\textbf{Micromagnetic simulations.}
To substantiate these theoretical insights, we perform micromagnetic simulations of the two-dimensional SAF-SkL using the GPU-accelerated solver MuMax3 \cite{VansteenkisteAA2014}, which numerically solves the Landau-Lifshitz Gilbert (LLG) equation \cite{Landau1935}:

 \begin{equation}\label{eq:2}
    \frac{\partial \mathbf{m}}{\partial t}=-\gamma\mathbf{m}\times {\mathbf{H}_{\mathrm{eff}}}+\alpha\mathbf{m}\times\frac{\partial\mathbf{m}}{\partial{t}},
\end{equation}
where $\mathbf{m}$ is the unit magnetization vector, $t$ is the time, $\gamma$ is the gyromagnetic ratio, and $\alpha$ is the dimensionless Gilbert damping constant. The effective field $\mathbf{H}_{\mathrm{eff}}$ is the functional derivative of the total magnetic energy $W$, $\mathbf{H}_{\mathrm{eff}}=-\frac{1}{\mu_{0}M_{s}} \frac{\delta W}{\delta \mathbf{m}}$, where $M_{s}$ is the saturation magnetization, and $\mu_0$ is the vacuum permeability. $W$ consists of the exchange energy $E_{\mathrm{ex}}$, the anisotropic energy $E_{\mathrm{an}}$, the Zeeman energy $ E_{\mathrm{Zee}}$, the interlayer antiferromagnetic exchange energy $ E_{\mathrm{af}}$ and the interfacial Dzyaloshinskii-Moriya interaction energy $E_{\mathrm{DMI}}$.

Magnons are excited by a sinc-function magnetic field:
\begin{equation}
\mathbf{H}(t) = H_\mathrm{0} \frac{\sin\left[2\pi f_\mathrm{c} (t - t_0)\right]}{2\pi f_\mathrm{c}(t - t_0)} \hat{\mathbf{x}},
\end{equation}
with the amplitude $H_0 = 10$ mT, $t_0 = 1$ ns, and the cutoff frequency $f_\mathrm{c} = 125$ GHz, applied to a central $40 \times 40$ nm$^2$ region for 40 ns [as indicated by the black shaded regions in Figs. \ref{FIG2}(a) and (b)]. The magnetization evolution is recorded every 4 ps, from which the band structure is obtained via Fourier transformation (FFT) \cite{Wanghao2007,BuessPRL2005}. The frequency resolution of the spectrum obtained is 0.025 GHz. Periodic boundary conditions were applied in the $x$ and $y$ directions for the band structure calculations, while open boundary conditions were used for simulating spin wave propagation and the frequency comb characteristics.

\FloatBarrier

\vbox{}
\section{Data availability}
\label{se:Data availability}
 The data that support the plots within this paper are available from the corresponding author upon reasonable request.

\vbox{}
\section{Code availability}
\label{se:Code availability}
The code that supports the plots within this paper is available from the corresponding author upon reasonable request.

\vbox{}
\section{References}
\label{se:References}


\vbox{}
\noindent\textbf{Acknowledgements}

\noindent
We thank Z. Jin and S. Lin for helpful discussions. This work was supported by the National Key R$\&$D Program under Contract No. 2022YFA1402802, the National Natural Science Foundation of China (NSFC Grants No. T2495212, No. 12074057, and No. 12374103), and Sichuan Science and Technology Program (No. 2025NSFJQ0045). X. L. acknowledges the support from the National Natural Science Foundation of China (Grant No. 12104322), Guangdong Basic and Applied Basic Research Foundation (Grant No. 2025A1515011895), and the Natural
Science Foundation of Top Talent of SZTU (Grant No. GDRC202309).

\vbox{}
\noindent\textbf{Author contributions}

\noindent
P. Y., X.J. L. and X. Z. conceived the idea. P. Y., C. Z. and H. L. coordinated and supervised the work. X.J. L. and Z. L. performed the micromagnetic simulation. X.J. L. and X. Z. carried out the theoretical analysis. X.J. L. and X. Z. drafted the study with the input from P. Y.. All the authors discussed the results and contributed to the manuscript. X.J. L. and X. Z. contributed equally to this work.

\vbox{}
\noindent\textbf{Competing interests}

\noindent
The authors declare no competing interests.
\end{document}